\documentclass{elsart}
\usepackage{graphicx,natbib}
\journal{New Astronomy}

\catcode`\à=\active \defà{\`a} \catcode`\À=\active \defÀ{\`A}
\catcode`\á=\active \defá{\'a} \catcode`\Á=\active \defÁ{\'A}
\catcode`\ä=\active \defä{\"a} \catcode`\Ä=\active \defÄ{\"A}
\catcode`\ã=\active \defã{\~a} \catcode`\Ã=\active \defÃ{\~A}
\catcode`\â=\active \defâ{\^a} \catcode`\Â=\active \defÂ{\^A}
\catcode`\å=\active \defå{{\aa}} \catcode`\Å=\active \defÅ{{\AA}}
\catcode`\ç=\active \defç{\c{c}} \catcode`\Ç=\active \defÇ{\c{C}}
\catcode`\è=\active \defè{\`e} \catcode`\È=\active \defÈ{\`E}
\catcode`\é=\active \defé{\'e} \catcode`\É=\active \defÉ{\'E}
\catcode`\ë=\active \defë{\"e} \catcode`\Ë=\active \defË{\"E}
\catcode`\ê=\active \defê{\^e} \catcode`\Ê=\active \defÊ{\^E}
\catcode`\ì=\active \defì{\`{\i}} \catcode`\Ì=\active \defÌ{\`{\I}}
\catcode`\í=\active \defí{\'{\i}} \catcode`\Í=\active \defÍ{\'{\I}}
\catcode`\ï=\active \defï{\"{\i}} \catcode`\Ï=\active \defÏ{\"{\I}}
\catcode`\î=\active \defî{\^{\i}} \catcode`\Î=\active \defÎ{\^{\I}}
\catcode`\ò=\active \defò{\`o} \catcode`\Ò=\active \defÒ{\`O}
\catcode`\ó=\active \defó{\'o} \catcode`\Ó=\active \defÓ{\'O}
\catcode`\ö=\active \defö{\"o} \catcode`\Ö=\active \defÖ{\"O}
\catcode`\ô=\active \defô{\^o} \catcode`\Ô=\active \defÔ{\^O}
\catcode`\õ=\active \defõ{\~o} \catcode`\Õ=\active \defÕ{\~O}
\catcode`\ù=\active \defù{\`u} \catcode`\Ù=\active \defÙ{\`U}
\catcode`\ú=\active \defú{\'u} \catcode`\Ú=\active \defÚ{\'U}
\catcode`\ü=\active \defü{\"u} \catcode`\Ü=\active \defÜ{\"U}
\catcode`\û=\active \defû{\^u} \catcode`\Û=\active \defÛ{\^U}
\catcode`\ý=\active \defý{\'y} \catcode`\Ý=\active \defÝ{\'Y}
\catcode`\ÿ=\active \defÿ{\"y} \catcode`\˜=\active \def˜{\"Y}
\catcode`\½=\active \def½{!`}
\catcode`\¾=\active \def¾{?`}
\catcode`\ß=\active \defß{{\ss}}

\begin{document}
\runauthor{Fonseca, Barbosa and Santos}
\begin{frontmatter}
\title{Site Evaluation and RFI spectrum measurements in Portugal at the frequency range 0.408-10 GHz for a GEM polarized galactic radio emission experiment}
\author[centra]{Rui Fonseca}
\author[centra,porto]{Domingos Barbosa\thanksref{emaild}}
\author[cfn]{Luis Cupido}
\author[aveiro1,aveiro2]{Dinis M. dos Santos}
\author[lbl,ucb]{George F.Smoot}
\author[inpe]{Camilo Tello}

\thanks[emaild]{email: barbosa@astro.up.pt}
\address[centra]{CENTRA, Instituto Superior Técnico, Av. Rovisco Pais, 1049-001 Lisbon, Portugal}
\address[aveiro1]{Dep. Electrónica e Telecomunicações, University of Aveiro, Campus Universitário de Santiago, 3810-193 Aveiro, Portugal}
\address[aveiro2]{Instituto de Telecomunicações, Campus Universitário de Santiago, 3810-193 Aveiro, Portugal}
\address[porto]{Centro de Astrofísica da Universidade do Porto, Rua das Estrelas, 4150-762 Porto, Portugal}
\address[lbl]{Astrophysics Group, MS 50-205, Lawrence Berkeley National Laboratory, 1 Cyclotron Rd., Berkeley CA 94720, USA}
\address[ucb]{Physics Dpt., University of California, 366 Le Conte Hall, Berkeley CA 94720, USA}     
\address[inpe]{Instituto Nacional de Pesquisas Espaciais, Divisão de Astrofísica, Caixa Postal 515, 12210-070, São José dos Campos SP, Brasil}
\address[cfn]{Centro de Fusão Nuclear, Instituto Superior Técnico, Av. Rovisco Pais, 1049-001 Lisboa, Portugal}
\begin{abstract}
We probed for Radio Frequency Interference (RFI) for the three potential Galactic 
Emission Mapping Experiment (GEM) sites at Portugal using custom made 
omnidirectional disconic antennas. For the installation of a 10-meter dish 
dedicated to the mapping of Polarized Galactic Emission foreground planned 
for 2005-2007 in the 5-10 GHz band, the three sites chosen as suitable to 
host the antenna were surveyed for local radio pollution in the frequency 
range [0.01-10] GHz. Tests were done to look for radio broadcasting and 
mobile phone emission lines in the radio spectrum. The results show one of 
the sites to be almost entirely RFI clean and showing good conditions to 
host the experiment. 
\end{abstract}
\end{frontmatter}

\section{Introduction} 

Cosmic Microwave Background (CMB) cosmology made a huge
leap forward from COBE maps\cite{smoot} towards high resolution map-making. The data returned from
recent and planed experiments (MAXIMA, Boomerang, DASI, NASA's WMAP and ESA's Planck
Surveyor satellite (launch 2007) offer a direct glimpse into the physics at
the surface of last scattering, providing constraints on cosmological
parameters and tests of theories of large scale structure formation and
favoring the inflationary paradigm. Obscuring our view of the CMB are
extragalactic and galactic foregrounds and the maximum cosmological
information can only be obtained if the foregrounds are optimally extracted.
Different physical components along the line of sight can be separated from
the underlying cosmic signal by using knowledge of their spectral an spatial
characteristics, given a sufficiently dense sampling in frequency and spatial
position as an use of prior knowledge. Typically diffuse galactic emission is
dominated by synchrotron radiation below 60GHz and by thermal dust emission
above 60GHz\cite{angelica}. Recently, CMB polarization is currently being detected (DASI -
2002, WMAP - 2003) see\cite{dasi1,dasi2,wmap1,wmap2} and will constitute for the next decade the best probe of
early Universe's physics.

Yet, while theoretically foreground amplitudes would be generically
distinguished by observing the different emission components where they are
dominant, there are no reasonably known templates accounting for their
amplitudes and effects. While for CMB total power (temperature) the foreground
amplitudes can be reasonably estimated, the amplitude of polarized foreground
like synchrotron or spinning dust is not known accurately. 

\subsection{Current GEM project}

To address the problem of foreground estimation, the Galactic Emission Mapping project started as an international collaboration ({\tt http://aether.lbl.gov} for detailed information) operating a portable 5.5-m dish with extensions to a 10-m surface capable of  
measuring the galactic emission at several latitudes in a wide range of 
frequencies \cite{torres}. To integrate for large sky areas and since sensitivity is more important than resolution, GEM scanning strategy consists 
on an slow azimuthal dish rotation until the required sensitivity is attained (see figure 1).
 The resultant maps obtained at several locations would 
then be merged to produce templates covering large areas of the sky with 
constant angular resolution from 408 MHz up to 10 GHz and good absolute calibration of the zero-level of the maps. 
\begin{figure*}
\begin{center}

\includegraphics[width=7cm]{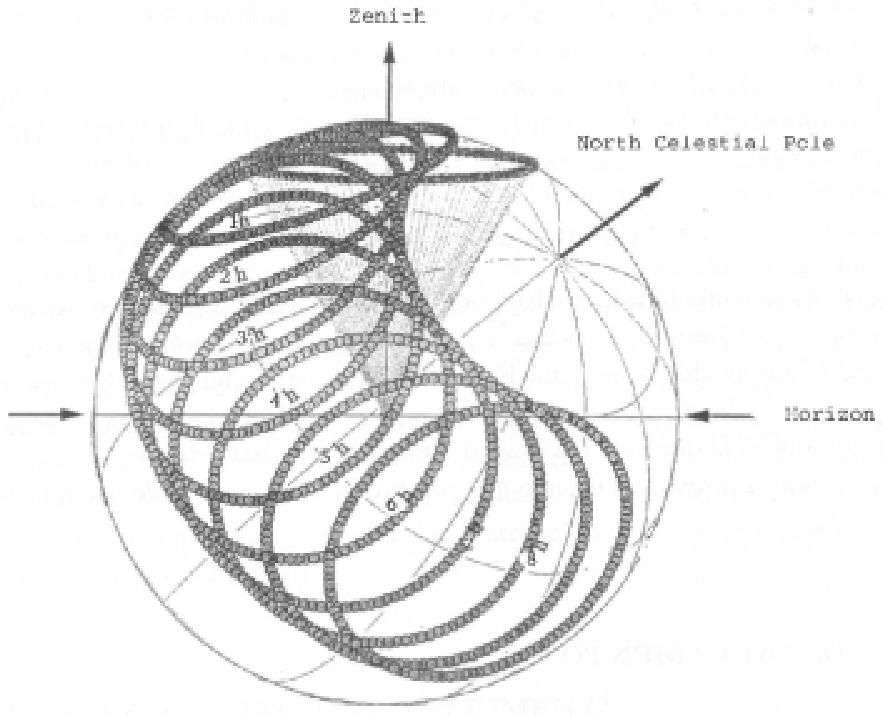}
\includegraphics[width=7cm]{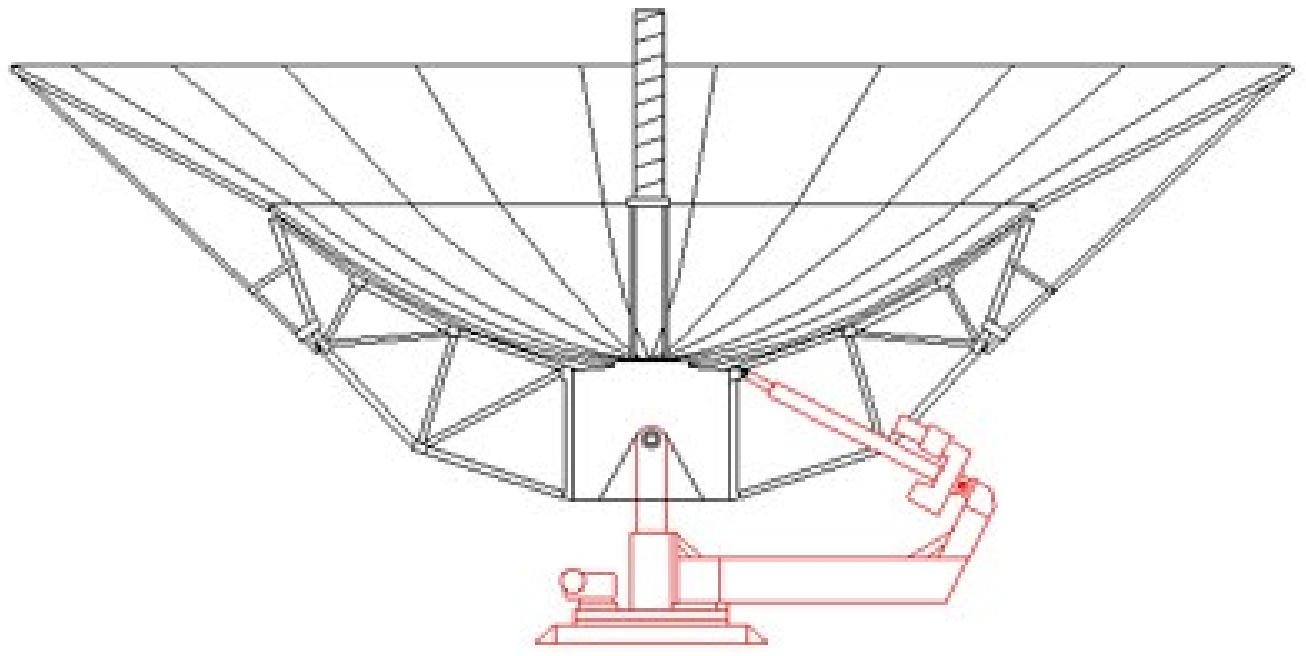}
\caption{Top figure represents a schematic view of the scanning strategy : 
azimuthal dish rotation for a given elevation. Bottom figure is a scheme of scanning instrument and mounting.}
\end{center}
\end{figure*}

With foreground cartography as one of
the main task currently being pursued by CMB teams, GEM project is evolving 
towards the measurements of galactic synchrotron polarization characterization at higher frequencies 5-10 GHz (Tello et al., in preparation). 
Originally, the Berkeley team has developed a compact and portable 5.5-m diameter radio antenna, which has been used for the first-stage observations. Observations were made from different locations, like near Bishop, California (fall 1993 through fall 1994), Leyva, Colombia (close to the equator 1995) and is currently working at Cachoeira Paulista, Brazil \cite{tello,tello1} having now 
coverd most of the southern hemisphere in 1.4 and 2.3 GHz. To also cover the
 northern hemisphere, and thus produce a template of most of the sky, one 
needs a good site in the northern hemisphere with suitable conditions for clima, RFI and infrastructure capabilities. Thus, Portugal was surveyed to find a suitable location to host a second antenna to complement the original GEM southern hemisphere maps. 
The GEM working frequencies were chosen because below 20 GHz, atmosphere contribution is negligible\cite{brandt}. Also, for polarization measurements, besides water lines around 22 GHz, oxygen only contaminates polarization measurements 
for frequencies higher than 60 GHz, where oxygen molecular rotational modes 
may be an important concern for ground experiments\cite{han,pardo}.

\subsection{RFI sources}

Radio observations need above all sites with no radio frequency interference (RFI).
Perhaps, the biggest threat lies on Global System for Mobile Communications
(GSM) networks, meaning frequency dispute with radio astronomers. GSM
mobile phone networks use several bands around 900 MHz, 1.8 GHz and the near
future Universal Mobile Telecommunications System (UMTS) to be shortly
implemented will use several 2.1 GHz bands. Other telecommunications
concerns are radio broadcast emissions, analog and digital television broadcast
Broadcast emissions, aeronautical communications - mainly along air corridors
and airports, satellite communications (communications, meteorological and
GPS) and amateur radio services, these using
theoretically a wide range of bands from 3.5 GHz to 250 GHz. Besides 
telecommunications, recent wireless computer networks use heavily
the 2.37 GHz band with prospects for a near future 5 GHz upgrade. Also,
microwave ovens will present leaks at 2.4 GHz despite the fact that they are 
in accordance with industrial standards. These leaks are insignificant to 
human health, but readily
detected as bursts of microwave signal in nearby radiometers working at these
frequencies. Besides these main concerns, secondary concerns as sources of
 RFI can be high-voltage power lines and old motorcycle engines. To conclude, 
one must avoid contact with human settlements. Of course, not all attributed 
frequencies by the telecommunications regulation 
authorities are susceptible of being contaminated. Either because they are 
used sporadically - easily eliminated in data processing - or because they are 
used heavily only in urban areas and do not appear in rural, isolated areas. 

\begin{table}
\caption{Typical Sources of RFI contaminations. Actually, while Radio Amateur 
service bands can go up to 250 GHz, they rarely exceed 1300 MHz.}
\label{RFsource}
\begin{center}
\begin{tabular}{|l | c|}
\hline
RF source & Frequency bands (MHz)\\ \hline
GSM networks & 890-960; 1710-1880 \\
UMTS networks & 1900-1980; 2110-2170\\
Radio broadcast  & $< 230$ \\
TV broadcast     &   475-870\\
Satellite commu. &   20000\\
Microwave Ovens  & 2450  \\
Computer Wireless Net. & 2370; 5800\\
Radio Amateur    & 0.177-1300 \\
\hline
\end{tabular}
\end{center}
\end{table}

\section{Sites evaluation} 

To find the best locations to host the experiment, several sites were selected 
after analyzing long term data of important weather variables - temperature, 
relative humidity, insolation rate (giving an idea of good weather days) from 
their geographic areas. The data statistics is publicly available from the 
Portuguese Instituto de Metereologia e Geofísica ({\tt http://www.meteo.pt}).
Portugal shows a temperate Atlantic climate with large variations despite its 
size. Main differences rest on altitude (mountainous and medium-high north/center and flat-low south) and humidity, depending on sea distance. Although 
Atlantic winds induce high levels of rain in the winter season and mean 
high levels of humidity nationwide except for some parts in the interior 
center and south, spring and summer mean relative humidities around 20-30\% 
or less  in the country's interior. These regions are also those with lower 
human density with some villages living still outside the GSM world. After 
correlating with GSM coverage maps, 
two sites in southern Portugal and one in central Portugal were therefore 
chosen for RFI measurements. 

The sites main characteristics show a similar 
pattern for annual RH variation of 20-30\% during most of the year. 
Temperatures tend to be high in summer($\sim 30-35^o$), with high thermal 
amplitudes between night and day 
specially for site 1 (average night temperature of $\sim 15^o$C or below). 
These values indicate good conditions for night observations throughout the 
dry season, 
implying low water absorption at the GEM considered frequencies \cite{butler}
Precise full characterization of sites weather conditions, specially of 
Castanheira da Serra, will be available locally after installing 
instrumentation for weather measurements in the exact places and should enable 
a robust check on variations of temperature and relative humidity. We plan to 
proceed with campaigns to check explicitly for the sites for annual, monthly, 
diurnal direct variations of water vapor content and temperature. 


\begin{table}

\label{sites}
\begin{center}
\begin{tabular}{|l|l|l|l|}
\hline
Location & longitude & latitude & elevation \\ \hline
Califórnia & $08^o$ 01' N & $37^o$ 18' W& 408 m\\
Castanheira da Serra & $07^o$ 52' N & $40^o$ 11' W& 839 m\\
\hline
\end{tabular}
\caption{Best sites coordinates.}
\end{center}
\end{table}

\subsection{RFI measurements}

After correlating both climatic and GSM service coverage, we selected 
several sites for survey. A primary search on GSM residual coverage reduced
our target sample to three sites. Since GEM scanning strategy involves 
azimuthal rotation of 360 degrees 
circles, the presence of even a single localized source of RFI signals can 
produce a substantial cut in the sky area surveyed. For this reason, we chose 
disconic antennas for its omnidirectionality and large band receiver 
capabilities(see figure2). While disconic antennas may theoretically work for bands with 
initial frequency 10 times larger than the central frequency (10:1), in 
practice we opted to optimally cover a frequency band range like 3:1. 
For the wide frequency range [100MHz-10 GHz] we built three disconic antennas
 covering optimally the bands [350-1050]MHz, [1.1-3.6]GHz, [3.6-10.8]GHz. The 
first antenna is, by above, still capable of detecting strong signals below 
100 MHz.

\begin{figure*}\label{fig_antena}
\begin{center}
\includegraphics[width=6cm]{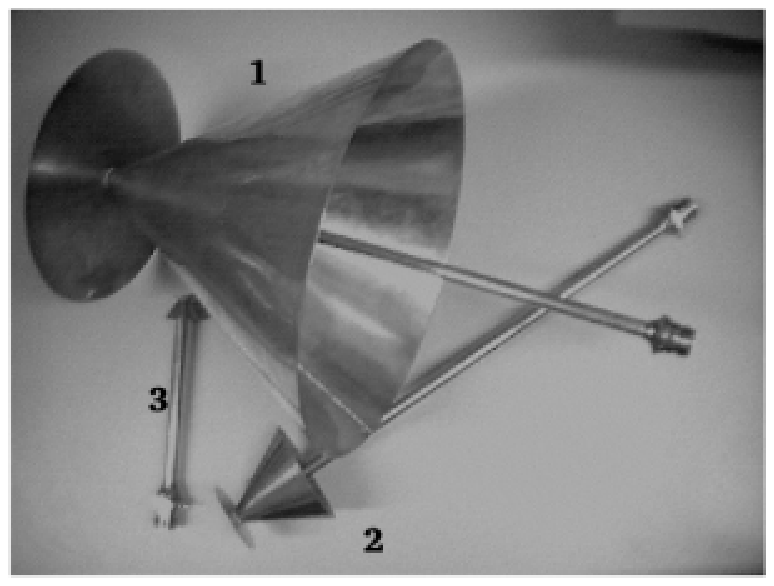}
\includegraphics[width=6cm]{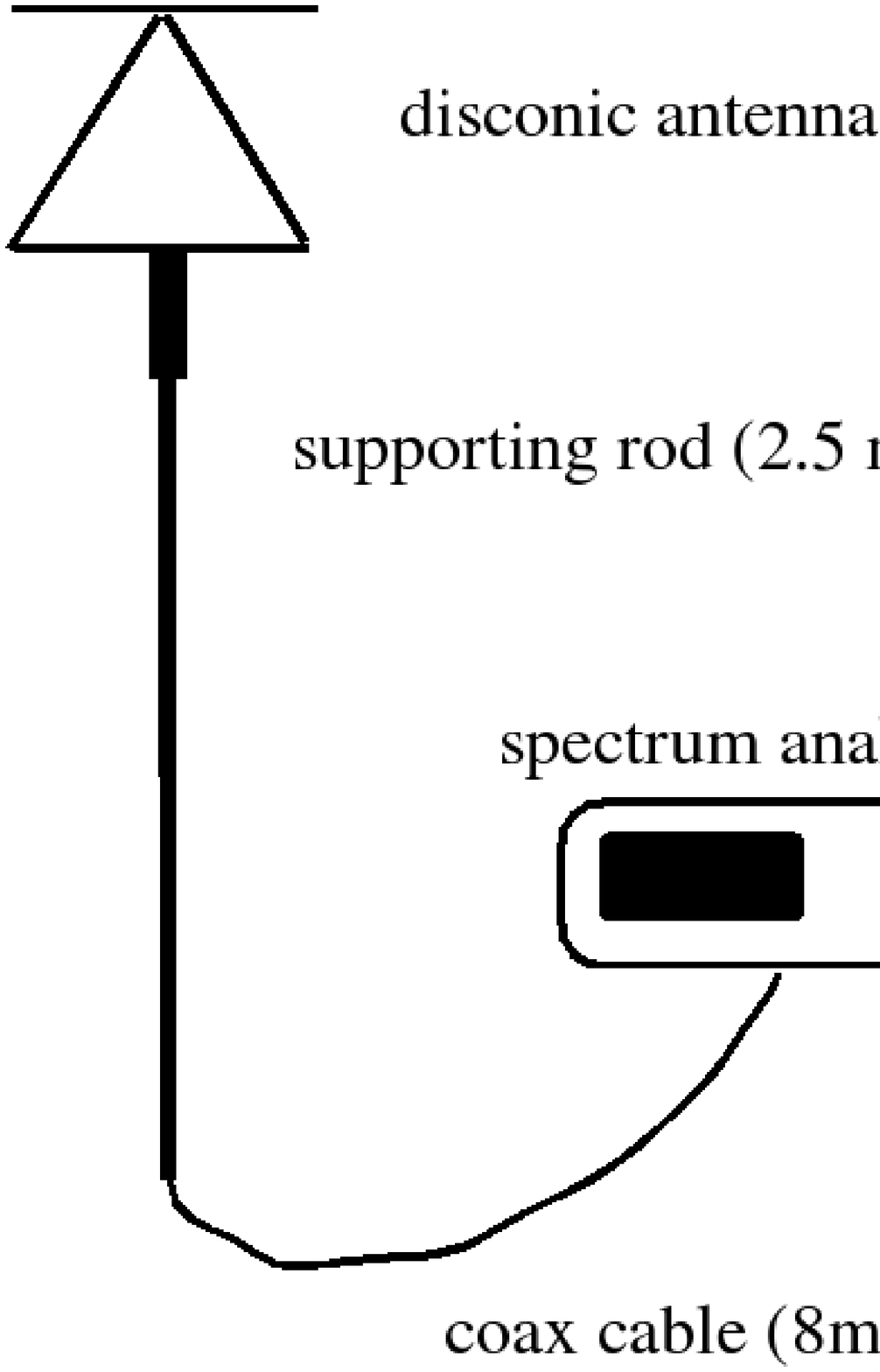}
\caption{The three disconic antennas used in our setup (right). They correspond to the bands:  1-[350-1050]GHz; 2-[1.1-3.6]GHz; 3-[3.6-10.8]GHz.}
\end{center}
\end{figure*}

The antennas 
were then put on a 2.5-m high rod and connected to a spectrum analyzer (an HP8563A). Measurements were taken at different time to check for source variability. The 
results are shown in figures 1,2. It is quite clear that RFI appears to be concentrated, 
as expected, in three wide bands (radio $\sim88-100$ MHz, tv $\sim500$ MHz 
and GSM 0.9-1.1 GHz.). Simultaneously, the relative humidity measured was 23\%.
The site of Castanheira da Serra, herafter site1, 
shows promising conditions with only one RFI source appearing intermittently 
(tv). While not showing any other expected source, it showed some very low 
frequency signals (shortwave signals), most likely due to ionospheric 
reflections. 
One may ask however, that weaker signals could be present and masked under our
 setup noise that could become important when a sensible receiver is setup on 
an antenna. 
 Ideally, the best situation would be to test for 
RFI with a very sensitive amplifier circuit as close as possible to our 
desired sensitivity. For 5-10 GHz, we expect the galactic emission to be on 
the order of 1-100 mK. The total power emitted at these frequencies, 
considering a 300 MHz bandwidth \cite{tello} is about -134 dBm\footnote{1 dBm is the unit for expression of power level in decibels with reference to a power of 1 milliwatt; dBm=$10\log(P_{mW})$ where $P_{mW}$ is the power in milliWatt.}, 
quite below our noise floor around -90 dBm. This shows that we should be aware of weak, 
distant signals that may be lurking below this preliminary survey noise sensitivity and 
carefully shield the instrument. We note that the expected dish will survey at elevations higher 
than $45^o$ to avoid horizon and ground problems, so sidelobes pick up would be much smaller. 
Typically a cassegrain antenna, optimized for low sidelobes, at an
elevation of $45^o$ to $60^o$ has an horizontal pick up lower than -40dB. 
A problem 
could be the very weak signals coming from an almost isotropic distribution of distant sources, leaking in our 
bands. However, human density (and village density) and as a consequence 
transmission antennas density in the areas we surveyed is very sparse. For site1, the nearest important settlement is at a distance of 15 Km, and there are 
several mountain rings in between. We note since we are going to operate near a protected 
radioastronomy band where no interference other than harmonics are expected, 
the main concern would be the intermodulation generated products 
intrinsic to our receiver front-end. Harmonics and spurious of the GSM base stations are obliged by law to be at least 60dB below carrier, therefore any unwanted signals that may fall inside our band would be at least 60 dB below the 
actuall received GSM signals. Also, RFI due to telecommunications may show strong 
variability with time and there could be the case where measurements were 
taken in a quiet period.  We did check for this and registered several 
intermittent signals, most likely due to the occasional use of GSM (figure2). 
In site1, the intermittent signal seems to be a distant TV broadcast leaking 
through. We did, however, check GSM and radio coverage from portuguese 
operators with the official portuguese frequency and radio communications 
board (ANACOM -{\tt http:www.anacom.pt}) and found 
the area of site1 to be a blank area (free of coverage). This 
was also checked directly at the nearest village, where GSM coverage is 
totally absent. Site1 also shows a very convenient orography with the presence
 of several higher mountains rings ($\sim1000$m) screening signals that could 
pass from nearby villages. 

Finally, to test our setup, we 
checked the antennas response, with the settings we used, at the laboratory 
(Instituto de Telecomunicações), to several generated signals and tested 
for GSM and weak wireless computer network signals down to the noise floor of -90 dBm.

\begin{figure*}
\begin{center}
\label{f2}
\includegraphics[angle=90]{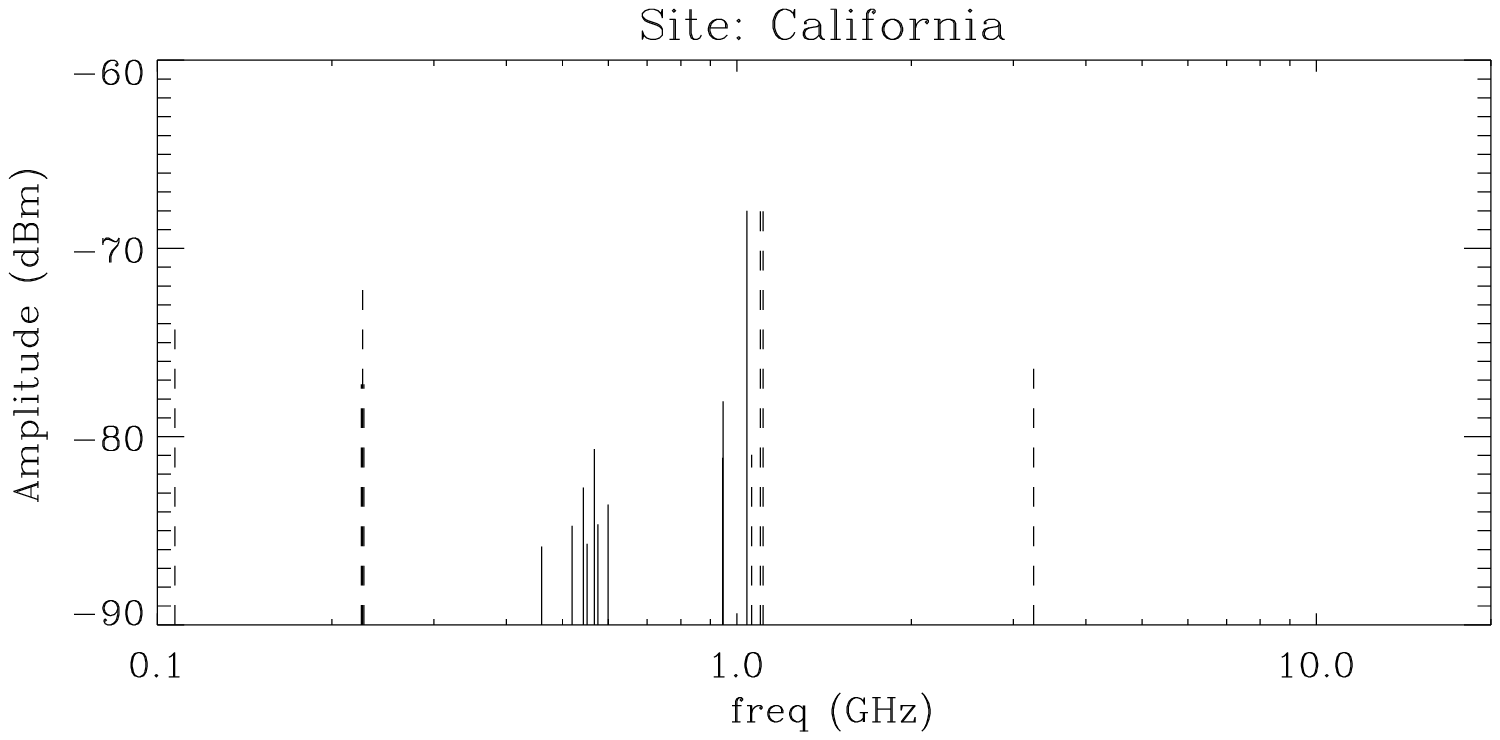}

\includegraphics[angle=90]{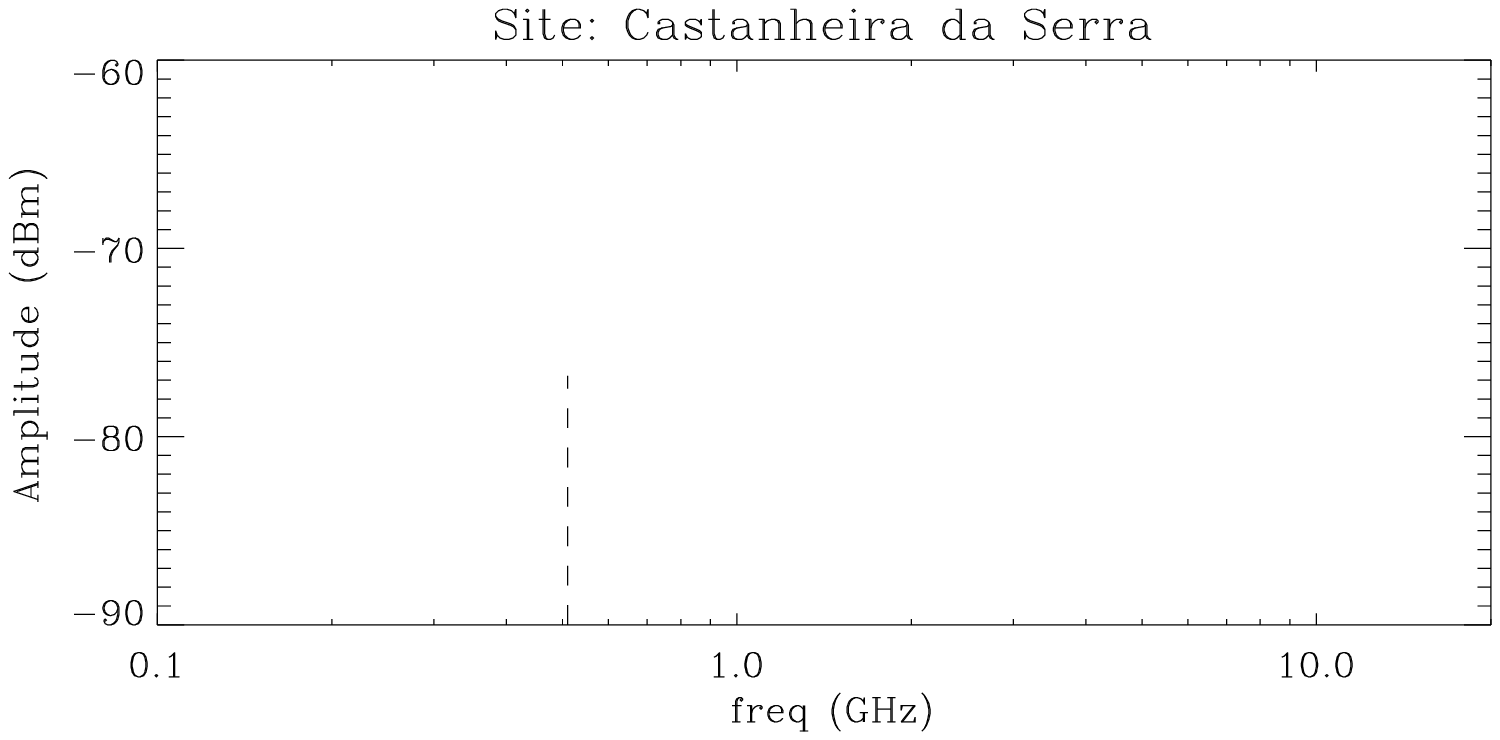}
\caption{RFI spectrum for the two sites. Signal amplitude are given in dBm. Solid lines are permanent signals and dashed lines represent intermittent signals detected during measurements. RFI lines appear, as expected by bunches: Radio services appears mainly at 100- 200 MHz; tv broadcast bunch around the 500 MHz band and GSM services clearly peak around the 900 MHz and the 1.1 GHz bands.}
\end{center}
\end{figure*}

\section{Conclusions} Within the context of the Galactic Emission project, we 
surveyed several sites for climatic and RFI measurements. Two of the sites 
show good conditions - low humidity, high number of good weather days, stable 
geograpy and low RFI on the survey bands to host a GEM antenna, with one - 
Castanheira da Serra clearly showing a clean radio spectrum, free of RFI in 
the important frequency range  of 2-10 GHz. 

\section{Acknowledgments}We would like to thank Juan Pardo 
for helpful discussions on atmosphere polarization. We acknowledge 
Miguel Lacerda and Mário Rui Santos at Instituto de Telecomunicações for all the help. We thank Ana Mourão for the encouragement and Prof. Armando Rocha for invaluable tips on antennas. DB acknowledges support from FCT through grant contract 
SFRH/11640/2002. This research was supported by FCT Project 
POCTI/FNU/42263/2001.

\end{document}